\documentstyle[epsf,twocolumn,prb,aps,psfig]{revtex}  

\begin{document}
\twocolumn[\hsize\textwidth\columnwidth\hsize\csname @twocolumnfalse\endcsname
\title{Universal superconducting and magnetic properties of the (Ca$_{x}$La$_{1-x}$)(Ba$_{1.75-x}$La$_{0.25+x}$)Cu$_{3}$O$_{y}$ system: a $\mu $SR investigation}
\author{$^{1}$Amit Keren, $^{1}$Amit Kanigel, $^{2}$James S. Lord, and $^{3}$Alex Amato }
\address{$^{1}$Physics Department, Technion-Israel Institute of Technology, Haifa 32000, Israel.\\
$^{2}$Rutherford Appleton Laboratory, Chilton Didcot, Oxfordshire OX11 0QX,
U.K.\\
$^{3}$Paul Scherrer Institute, CH 5232 Villigen PSI, Switzerland}
\date{\today}
\maketitle

\begin{abstract}
The (Ca$_{x}$La$_{1-x}$)(Ba$_{1.75-x}$La$_{0.25+x}$)Cu$_{3}$O$_{y}$ 
system is ideal for testing theories of high temperature superconductivity,
since nearly the full range of doping is controlled by $y$, and $T_{c}^{\max
}$ is continuously controlled by $x$, with minimal structural changes. We
investigate this system with both transverse and longitudinal field $\mu $
SR. This allows us to re-examine the Uemura relation, the nature of the
spontaneous magnetic fields below $T_{c}$, and the relation between their
appearance temperature $T_{g}$ and $T_{c}^{\max }$. Our major findings are:
(I) the Uemura relation is respected by CLBLCO more adequately than by other
cuprates, (II) $T_{g}$ and $T_{c}$ are controlled by the same energy scale,
(III) the phase separation between hole poor and hole rich regions is a
{\em microscopic} one, and (IV) spontaneous magnetic fields appear gradually with
no moment size evolution.
\end{abstract}
\vspace{0.5cm}
]

The Uemura relation is a milestone in the research of high temperature
superconductors [HTSC] and a key ingredient in most modern theories of HTSC.
This relation \cite{UemuraPlot} states that in underdoped HTSC the
superconductivity transition temperature $T_{c}$ is proportional to the muon
spin rotation [$\mu $SR] line width $\sigma $, and that the proportionality
constant is universal for all HTSC materials. Since $\sigma $ is
proportional to $\lambda _{ab}^{-2}$, where $\lambda _{ab}$ is the in-plane
penetration depth [see Sec.~\ref{TFmSR}], there is a one to one relation
between $T_{c}$ and $\lambda _{ab}^{-2}$ for all underdoped HTSC. However,
this relation seems to break in the optimally and over doped regions. To
date, the origin of this breakdown is not clear.

Another key ingredient of \ modern theories of HTSC is that at low
temperatures (T), cuprates phase-separate into regions that are hole
``poor'' and hole ``rich'' \cite{Kapitulnik1}. While hole rich regions
become superconducting below $T_{c}$, the behavior of hole poor regions at
these temperatures is not quite clear. Some data support the existence of
magnetic moments in these regions \cite{muonTg}, but there are still many
open questions regarding these moments and the spontaneous magnetic fields
associated with them. For example: Is there a true phase transition at $%
T_{g} $? What is the field profile and how is it different from, or similar
to, a canonical spin glass? Is the field confined solely to the hole poor
regions or does it penetrate the hole rich regions? Also, the interplay
between magnetism and superconductivity is not clear. Is strong magnetic
background beneficial or detrimental to superconductivity?

\begin{figure}[tbp]
\centerline{\epsfxsize=8.0cm \epsfbox{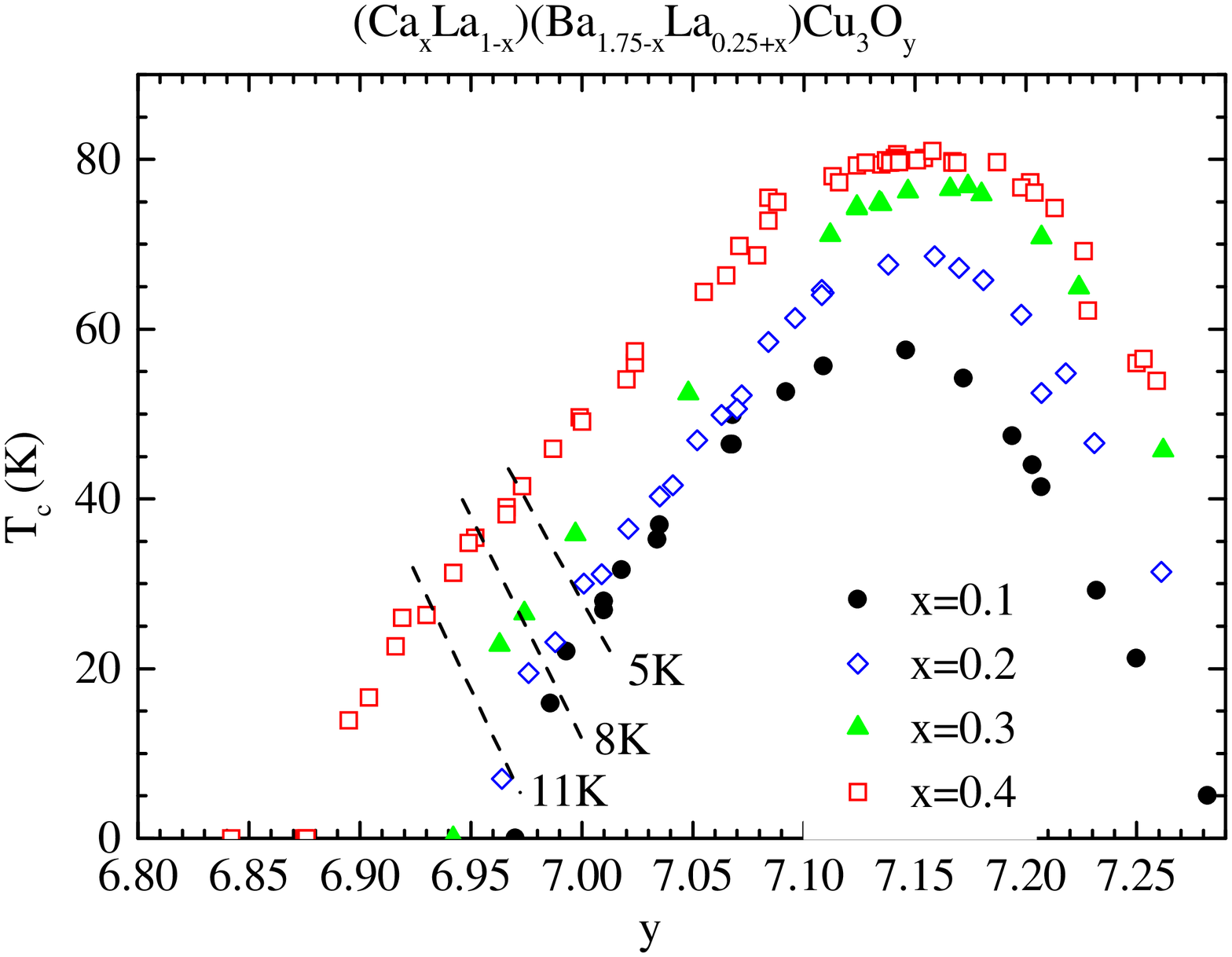}}                   
\caption[]{Phase diagram of 
(Ca$_{x}$La$_{1-x}$)(Ba$_{1.75-x}$La$_{0.25+x}$)Cu$_{3}$O$_{y}$. The dashed lines
indicate samples with equal $T_{g}$.}
\label{clblco}
\end{figure}

We address these questions by investigating the (Ca$_{x}$La$_{1-x}$)(Ba$%
_{1.75-x}$La$_{0.25+x}$)Cu$_{3}$O$_{y}$ (CLBLCO) family of superconductors.
These superconductors belong to the 1:2:3 family and has several properties
that make it ideal for our purpose. It is tetragonal throughout its range of
existence $0\leq x < 0.5$, so there is no ordering of CuO chains.
Simple valence sums \cite{comment3}, more sophisticated bond-valance
calculations \cite{Eckstein1}, and thermoelectric power measurements \cite
{Arkady1} show that the hole concentration is $x$ independent. As shown in
Fig.~\ref{clblco}, by changing $y$, for a constant value of $x$, the full
superconductivity curve, from the underdoped to the overdoped, can be
obtained. Finally, for different Ca contents, parallel curves of $T_{c}$ vs $%
y$ are generated. Therefore, with CLBLCO one can move continuously, and with
minimal structural changes, from a superconductor resembling YBCO to one
similar to LSCO. We study the superconducting and magnetic properties of
CLBLCO by performing, zero, longitudinal, and transverse field muon spin
relaxation experiments. The report on the zero and longitudinal field
measurements is an extension of our recent letter \cite{KanigelPRL02}. 
\section{Experimental aspects}

The preparation of the samples is described elsewhere \cite{clblco1}. Oxygen
content was determined using iodometric titration. All the samples were
characterized using X-ray diffraction and were found to be single phase. $%
T_{c}$ presented in the phase diagram is obtained from resistivity
measurements.

\begin{figure}[tbp]
\centerline{\epsfxsize=8.0cm \epsfbox{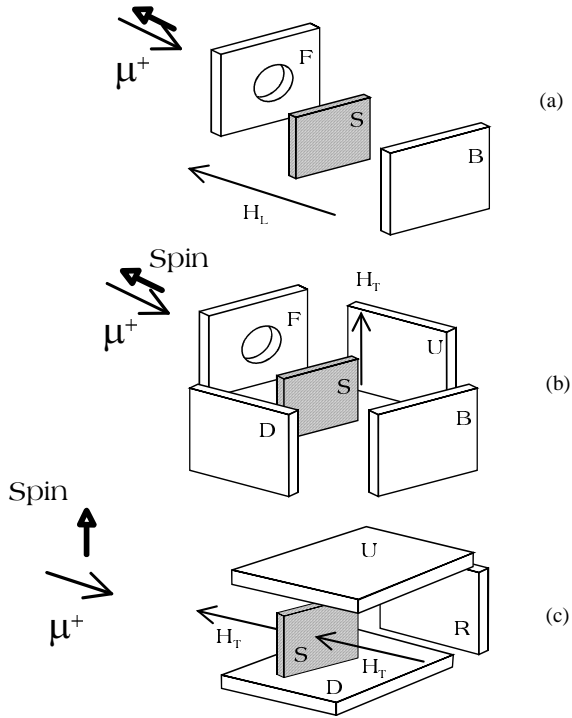}}                   
\caption[]{$\protect%
\mu $SR experimental setup representing the (a) longitudinal and transverse
field configurations in the (b) ISIS and (c) PSI facilities.  }
\label{MuSRSetup}
\end{figure}

Our $\mu $SR experiments were done at two facilities. When a good
determination of the base line was needed we used the ISIS pulsed muon
facility at Rutherford Appleton Laboratory, UK. When high timing resolution
was required we worked at the Paul Scherrer Institute, Switzerland (PSI). In 
$\mu $SR experiments one injects polarized muons into the sample and applies
a magnetic field. Decay positrons are emitted preferentially in the
direction parallel to the muon spin and are detected by positron counters.
We used two types of setups, the transverse field [TF] and the longitudinal
field [LF], which also includes zero field, as shown in Fig.~\ref{MuSRSetup}%
. The LF configuration works as in panel (a) in both facilities. The TF
configurations work in ISIS as in panel (b) and at PSI as in panel (c). From
the counters an histogram is generated of positrons counts as a function of
time, where the time is measured from the moment the muons enter the sample.
The asymmetries $A_{FB}(t)$, $A_{UD}(t)$ and $A_{LR}(t)$ are than generated
by $A_{FB}(t)=\left[ F(t)-B(t)\right] /\left[ F(t)+B(t)\right] $ and
similarly with $U$, $D$ instead of $F$, $B$, where $F$, $B$, $U$, $D$ are
the forward, backward, up, \ and down counters. $A_{LR}(t)$ is generated
only from the left counter in Fig.~\ref{MuSRSetup}(c). Most of the data were
taken with a $^{4}$He cryostat. However, in order to study the internal
field profile we had to avoid dynamical fluctuations by freezing the moments
completely. For this purpose we used the $^{3}$He cryostat at ISIS with a
base temperature of $350$~mK. All $\mu $SR measurements were done on
sintered pellets.

\section{TF-$\protect\mu $SR}
\label{TFmSR}

\begin{figure}[tbp]
\centerline{\epsfxsize=8.0cm \epsfbox{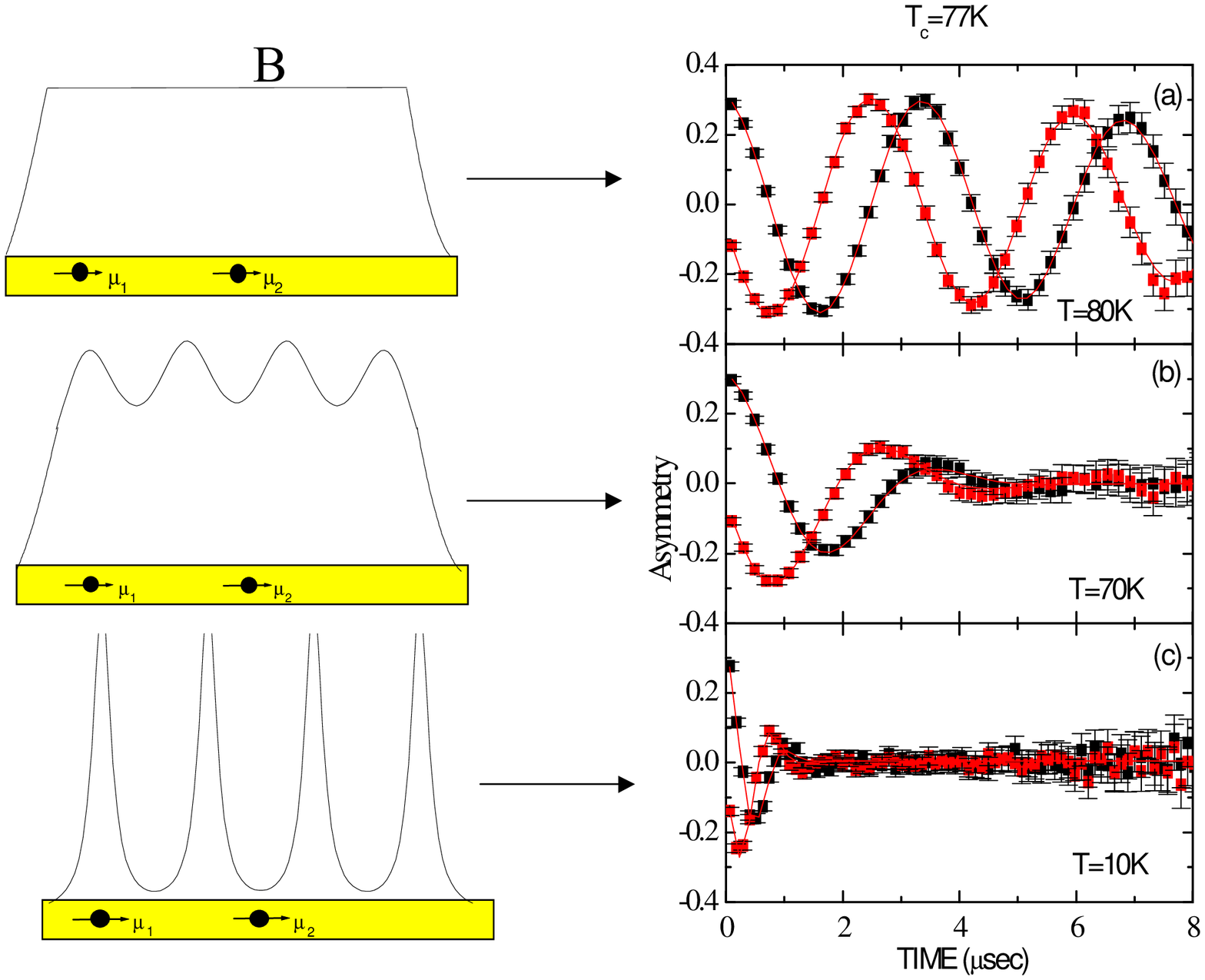}}       
\caption[]{Demonstrating the relation
between the field distribution and the real and imaginary asymmetries in a
TF-$\protect\mu $SR experiment. This data was taken at ISIS and is presented
in a rotating reference frame.}
\label{RlxDemo}
\end{figure}

These experiments are done by field cooling (FC) the sample to 1.8~K at an
external field of 3~kOe in PSI and 400~Oe in ISIS. As explained above we
apply the field perpendicular to their spin direction, and every muon then
precesses according to the local field in its environment. When field
cooling the sample, a vortex lattice is formed, and the field from these
vortices decay on a length scale of $\lambda $. This leads to a
inhomogeneous field distribution in the sample. Since the magnetic length
scale is much larger than the atomic one, the muons probe the magnetic field
distribution randomly, which, in turn, leads to a damping of the muons
average spin polarization. This situation is demonstrated in Fig.~\ref
{RlxDemo} where we present an image of the field profile, and the
corresponding the real and imaginary part of the muon asymmetry. At
temperatures above $T_{c}$ the field is homogeneous and all muons experience
the same field, and therefore no relaxation is observed. Well below $T_{c}$
there are strong field variations and therefore different muon precess with
different frequencies, and the average polarization quickly decays to zero.
In intermediate temperatures the field variation are not severe and the
relaxation is moderate.

It was shown that in powder samples of HTSC the muon asymmetry $A(t)$ is
well described by \cite{MusrBook}, 
\begin{equation}
A(t)=\exp (-\sigma ^{2}t^{2}/2)\cos (\omega t+\varphi )  \label{TFFitFunc}
\end{equation}
where $\omega =\gamma _{\mu }H$ is the precession frequency of the muon, $%
\sigma $ is the relaxation rate, and $\varphi $ is a phase which depends on
the counters used to generate the asymmetry. Our analysis for both ISIS and
PSI data is done in a reference frame rotating at $\omega _{rrf}$ and the
real and imaginary components of the signal are fitted simultaneously.
Therefore, the frequency in Fig.~\ref{RlxDemo} is $\gamma _{\mu
}H-\omega _{rrf}$ where $\omega _{rrf}$ is chosen arbitrarily for
presentation purpose. The solid line in this figure is the fit result. The
fact that the whole asymmetry relaxes indicates that CLBLCO is a bulk
superconductor.

\begin{figure}[tbp]
   \centerline{\epsfxsize=8.0cm \epsfbox{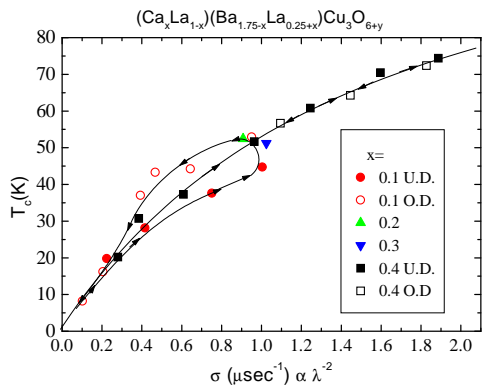}}                     
\caption[]{A Uemura plot showing $T_{c}$ vs.
the muon relaxation rate $\protect\sigma $ in Eq.~\ref{TFFitFunc} for the
CLBLCO family of superconductors.}
\label{UemuraPlot}
\end{figure}

The fit results for $\sigma $ are shown in Fig \ref{UemuraPlot}. As can be
seen, the dependence of $T_{c}$ on $\sigma $ is linear in the under-doped
region and universal for all CLBLCO families, as expected from the Uemura
relations. However, there is a new aspect in this plot. There is no
``boomerang'' effect, namely, overdoped and underdoped samples with equal $%
T_{c}$ have the same $\sigma $, with only slight deviations for the $x=0.1$
sample as demonstrated by the arrows in Fig.~\ref{UemuraPlot}. Therefore, in
CLBLCO there is one to one correspondence between $T_{c}$ and $\sigma $, and
therefore $\lambda _{ab}^{-2}$, over the whole doping range. This is our
first important finding.

\section{ZF-$\protect\mu $SR}

\begin{figure}[tbp]
      \centerline{\epsfxsize=8.0cm \epsfbox{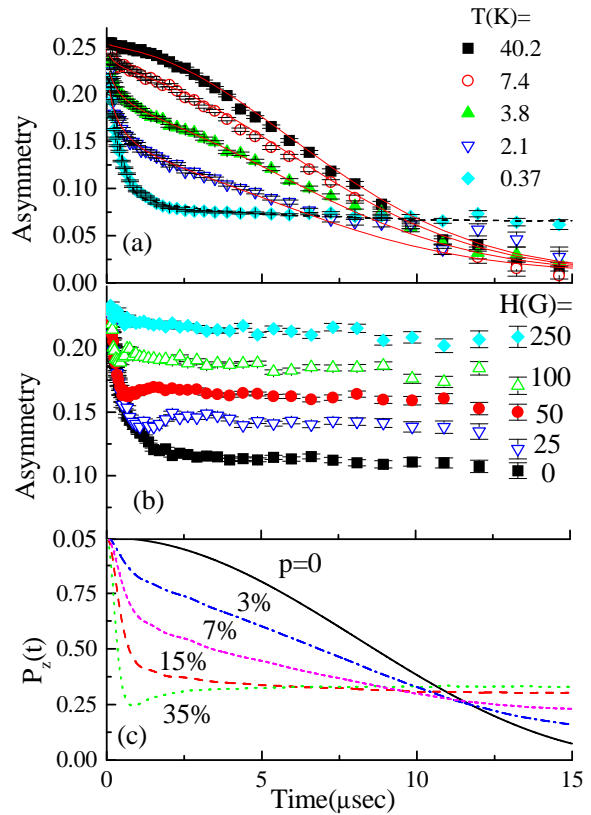}}               
\caption[]{(a) ZF-$\protect\mu $SR
spectra obtained in a $x=0.1$, $y=7.012$ sample with T$_{c}$=33.1K. The solid lines are fit to
the data using Eq.~\ref{func}, the dashed line is a fit using the simulation
as described in the text. (b) $\protect\mu $SR spectra obtained in
longitudinal fields from the $x=0.4$, $y=6.984$ sample at 350~mK. (c)
Polarization curves generated by the simulation program as described in the
text.}
\label{pol}
\end{figure}

Typical muon asymmetry depolarization curves are shown in Fig.~\ref{pol} (a)
for different temperatures in the $x=0.1$ and $y=7.012$ ($T_{c}=33.1K$)
sample. The change of the polarization shape with temperature indicates a
freezing process, and the data can be divided into three temperature
regions. In region (I), given by $T\gtrsim 8$~K, the muon relaxes according
to the well known Kubo-Toyabe (KT) function given by 
\begin{equation}
KT(t)=\frac{1}{3}+\frac{2}{3}(1-\Delta ^{2}t^{2})\exp (-\frac{1}{2}\Delta
^{2}t^{2}),  \label{KT}
\end{equation}
(see Eq.~\ref{rhotoP}) typical of the case where only frozen nuclear moments
are present \cite{MusrBook}. In region (II), bounded by $8$~K $\gtrsim
T\gtrsim 3$~K, part of the polarization relaxes fast and the rest relaxes as
in the first region. As the temperature is lowered the fast portion
increases at the expense of the slow one. Moreover, the relaxation rate in
the fast portion seems independent of temperature. Finally, at long time the
asymmetry relaxes to zero. In region (III), where $3$ K $\gtrsim T$, the
asymmetry at long times no longer relaxes to zero, but instead recovers to a
finite value. This value is $\simeq 1/3$ of the initial asymmetry $A_{z}(0)$.

To demonstrate that the internal field is static at base temperature, the
muon polarization was measured with an external field applied parallel to
the initial muon spin-polarization. This geometry allows one to distinguish
between dynamic and static internal fields. In the dynamic case the
asymmetry is field independent \cite{Comment1}. In contrast, in the static
case the total field experienced by the muon is a vector sum of $H$ and the
internal fields, which are of order $\left\langle B^{2}\right\rangle ^{1/2}$%
. For $H\gg \left\langle B^{2}\right\rangle ^{1/2}$ the total field is
nearly parallel to the polarization. Therefore, in the static case, as $H$
increases, the depolarization decreases, and the asymmetry recovers to its
initial value. Because we are dealing with a superconductor, this field
sweep was done in field-cool conditions. Every time the field was changed
the sample was warmed to above $T_{c}$ and cooled down in a new field. The
results are shown in Fig.~\ref{pol}(b). At an external field of 250~G, the
total asymmetry is nearly recovered. Considering the fact that the internal
field is smaller than the external one due to the Meissner effect, this
recovery indicates that the internal field is static and of the order of
tens of Gauss. Next we perform quantitative data analysis in two parts: high
temperatures (region II), and base temperature.

\subsection{High T Data Analysis}

First we discuss region II. Here we focus on the determination of $T_{g}$.
For that purpose we fit a combination of a fast relaxing function and a KT
function to the data \cite{Savici} 
\begin{equation}
A_{z}(t)=A_{m}\exp \left( -\sqrt{\lambda t}\right) +A_{n}KT(t),  \label{func}
\end{equation}
where $A_{m}$ denotes the amplitude of the magnetic part, $\lambda $ is the
relaxation rate of the magnetic part, and $A_{n}$ is the amplitude of the
nuclear part. The relaxation rate of the KT part was determined at high
temperatures and is assumed to be temperature independent. The sum $%
A_{m}+A_{n}$ is constrained to be equal to the total initial asymmetry at
high temperatures. The relaxation rate $\lambda $ is common to all
temperatures. The solid lines in Fig.~\ref{pol}(a) are the fits to the data
using Eq.~\ref{func}.

\begin{figure}[tbp]
\centerline{\epsfxsize=8.0cm \epsfbox{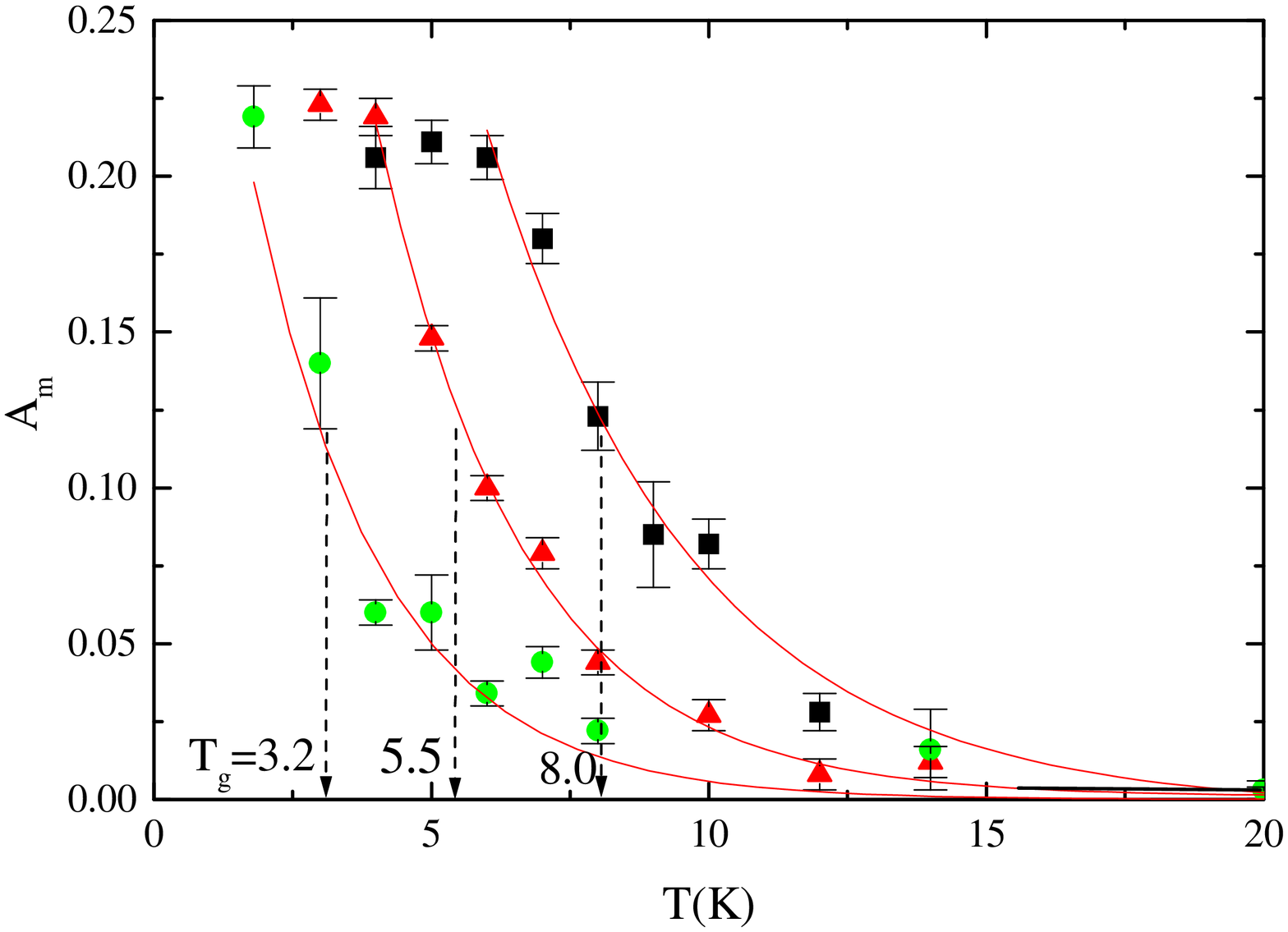}}                     
\caption[]{Magnetic amplitude as function of temperature f
or different samples. The solid lines are guides to the eye.}
\label{TgDetermination}
\end{figure}

The success of this fit indicates the simultaneous presence of two phases in
the sample; part of the muons probe the magnetic phase while others probe
only nuclear moments. As the temperature decreases $A_{m}$, which is
presented in Fig.~\ref{TgDetermination} for three samples, grows at the
expense of $A_{n}$. At low temperatures $A_{m}$ saturates to the full muon
asymmetry. A similar temperature dependence of $A_{m}$ is found in all
samples. The origin of the magnetic phase is electronic moments that slow
down and freeze in a random orientation. The fact that $\lambda $ is
temperature independent means that in the magnetic phase $\gamma _{\mu
}\left\langle B^{2}\right\rangle ^{1/2}$, where $\gamma _{\mu }$ is the muon
gyromagnetic ratio, is temperature independent. In other words, as the
temperature is lowered, more and more parts of the sample become magnetic,
but the moments in these parts saturate upon freezing.

\begin{figure}[tbp]
\centerline{\epsfxsize=8.0cm \epsfbox{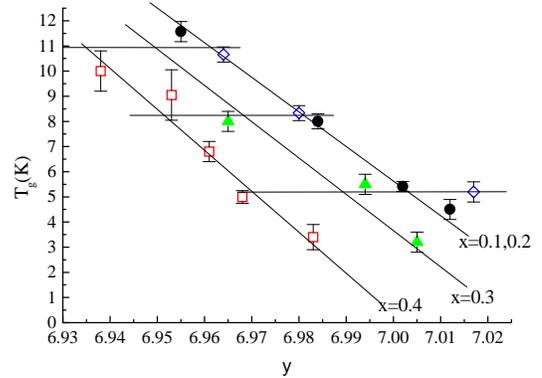}}                     
\caption[]{$T_{g}$ vs. y. The horizontal solid
lines are the equal $T_{g}$ lines appearing in Fig.~\ref{clblco}.}
\label{Tg}
\end{figure}

Our criterion for $T_{g}$ is the temperature at which $A_{m}$ is half of the
total muon polarization as demonstrated by the vertical lines in Fig.~\ref
{TgDetermination} for the three different samples. The phase diagram that is
shown in Fig.~\ref{Tg} represents $T_{g}$ for various samples differing in
Ca and O contents. This diagram is systematic and rather smooth suggesting
good control of sample preparation. As expected, for constant $x$, higher
doping gives lower $T_{g}$.

We have singled out three groups of samples with a common $T_{g}=11$, $8$
and $5$~K as shown in Fig.~\ref{Tg} by the horizontal solid lines. These
samples are represented in the phase diagram in Fig.~\ref{clblco} by the
dotted lines. The phase diagram, containing both $T_{g}$ and $T_{c}$, is the
second main finding of this work. It provides clear evidence of the
important role of the magnetic interactions in high temperature
superconductivity as discussed in Sec.~\ref{Discussion}.

\subsection{Low T Data Analysis}

\begin{figure}[tbp]
   \centerline{\epsfxsize=8.0cm \epsfbox{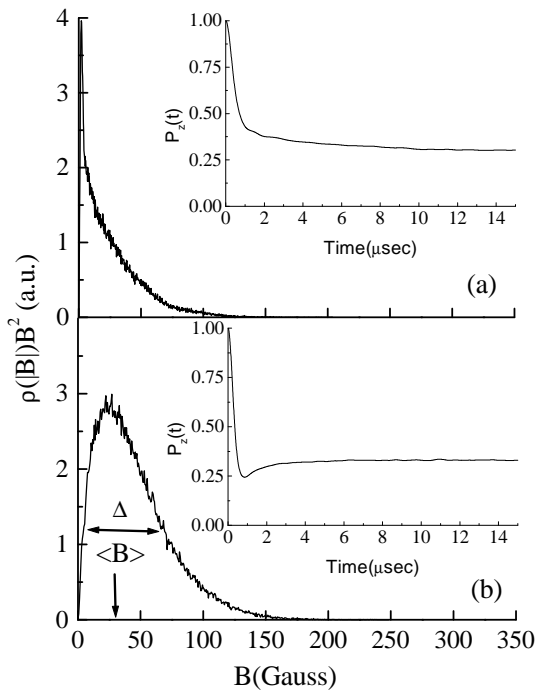}}                  
\caption[]{(a) The internal field
distribution extracted from the simulations for the case of correlation
length $\protect\xi =3$ lattice constants, maximum moment size of $0.06%
\protect\mu _{B}$ and magnetic moment concentration $p=15\%$. Inset: The
muon spin polarization for that distribution. (b) The same as above for the
case of $p=35\%$.}
\label{model}
\end{figure}

\begin{figure}[tbp]
\centerline{\epsfxsize=8.0cm \epsfbox{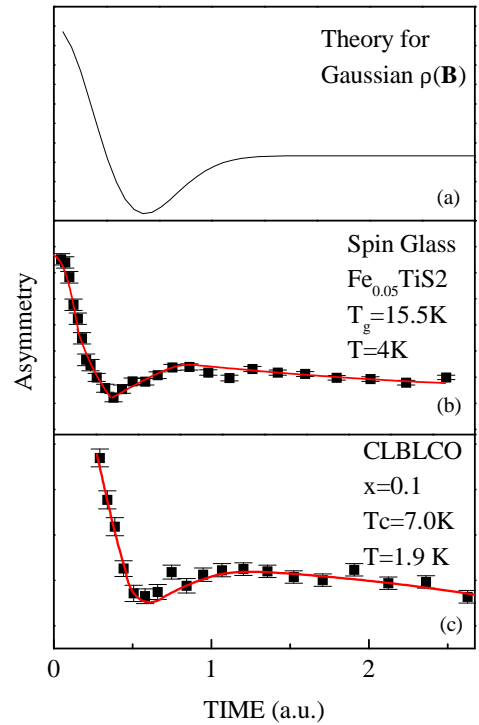}}                     
\caption[]{Demonstrating the expected muon
spin plarization function for (a) a Gaussina field distribution, (b) in a
canonical spin glass Fe$_{0.05}$TiS$_{2}$, and (c) extremely underdoped
CLBLCO.}
\label{DeepDemon}
\end{figure}

We now turn to discuss the muon depolarization at base temperature. In this
case all the muons experience only a static magnetic field, as proven above.
This allows one to reconstruct the internal field distribution out of the
polarization curve. The polarization of a muon spin experiencing a unique
field ${\bf B}$ is given by $P_{z}(t)=\cos ^{2}(\theta )+\sin ^{2}(\theta
)\cos (\gamma |{\bf B}|t)$, where $\theta $ is the angle between the field
and the initial spin direction. When there is an isotropic distribution of
fields, a 3D powder averaging leads to 
\begin{equation}
P_{z}(t)=\frac{1}{3}+\frac{2}{3}\int_{0}^{\infty }\rho (|B|)\cos (\gamma
|B|t)B^{2}dB  \label{rhotoP}
\end{equation}
where $\rho (|B|)$ is the distribution of $\left| {\bf B}\right| $.
Therefore, the polarization is given by the Fourier transform of $\rho
(|B|)B^{2}$ and has a $1/3$ base line. When the distribution of ${\bf B}$ is
centered around zero field, $\rho (|B|)B^{2}$ is a function with a peak at $%
\langle B\rangle $ and a width $\Delta $, and both these numbers are of the
same order of magnitude [e.g. Fig~\ref{model}(b)]. Therefore we expect the
polarization to have a damped oscillation and to recover to $1/3$, a
phenomenon known as the dip [e.g. the inset in Fig~\ref{model} (b)].
Gaussian, Lorentzian and even exponential random field distribution \cite
{Larkin}, and, more importantly, all known canonical spin glasses, produce
polarization curves that have a dip before the $1/3$ recovery. This is
demonstrated in Fig.~\ref{DeepDemon}. For a Gaussian distribution of width $%
\Delta $ we obtain Eq.~\ref{rhotoP} which is demonstrated in panel (a). The
cases of a canonical spin glass Fe$_{0.05}$TiS$_{2}$, and an extremely
underdoped CLBLCO are presented in panels (b) and (c). Furthermore, a
dipless polarization curve that saturates to $1/3$ cannot be explained using
dynamical arguments. Therefore, the most outstanding feature of the muon
polarization curve at base temperature is the fact that no dip is present,
although there is a $1/3$ tail. This behavior was found in all of our
samples with $T_{c}>7$~K, and also in Ca doped YBCO \cite{Nieder2} and Li
doped YBCO \cite{Mendels_Private}.

The lack of the dip in $P_{z}(t)$ can tell much about the internal field
distribution. It means that $\langle B\rangle $ is much smaller than $\Delta 
$. In that case the oscillations will be over-damped and the polarization
dipless! In Fig.~\ref{model} we show, in addition to the $\langle B\rangle
\simeq \Delta $ case described above [panel (b)], a field distribution that
peaks around zero [panel (a)]. Here $\langle B\rangle $ is smaller than $%
\Delta $, and, indeed, the associated polarization in the inset is dipless.
Thus in order to fit the base temperature polarization curve we should look
for $\rho (|B|)B^{2}$ with most of its weight around zero field. This means
that $\rho (|B|)$ diverges like $1/B^{2}$ at $|B|\rightarrow 0$, namely,
there is abnormally high number of low field sites.

It also means that the phase separation is not a macroscopic one. If it
were, all muons in the field free part would probe only nuclear moment and
their polarization curve should have a dip or at least its beginning as in
the high temperature data. The same would apply for the total polarization
curve, in contrast to observation. Thus, the superconducting and magnetic
regions are intercalated on a microscopic scale ($\sim 20\AA $) \cite
{comment2}. This is the third main finding of this work.

The special internal field distribution, and the nature of the gradual
freezing of the spins, can be explained by the intrinsic inhomogeneity of
hole concentration. The part of the sample that is hole poor, and for that
reason is ``more'' antiferromagnetic, will freeze, while the part which is
hole rich will not freeze at all. The variation in the freezing temperature
of different parts of the sample can be explained by the distribution of
sizes and hole concentration in these antiferromagnetic islands \cite{Cho}.
The large number of low field sites is a result of the fact that the
magnetic field generated in the magnetic regions will penetrate into the
hole rich regions but not completely.

\section{Numerical simulation}

To improve our understanding of the muon polarization, we performed
simulations of a toy model aimed at reproducing the results described above.

\begin{figure}[tbp]
 \centerline{\epsfxsize=8.0cm \epsfbox{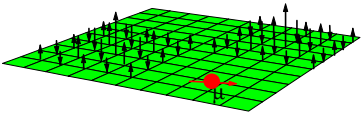}}                    
\caption[]{Demonstrating the numerical simulations.
Two spins (long arrows) are placed on the lattice. They polarize the near by
spins. The muon interacts with the spin by dipolar interaction. Nuclear
moment (which participate in the simulation) are not shown.}
\label{Simulation}
\end{figure}

A 2D $100\times 100$ square lattice is filled with two kinds of moments,
nuclear and electronic. All the nuclear moments are of the same size, they
are frozen and they point in random directions. Of the electronic moments
only a small fraction $p$ is assumed to be frozen; they represent magnetic
regions with uncompensated antiferromagnetic interactions. Since these
regions may vary in size, the moments representing them are random, up to a
maximum size. The frozen electronic moments induce spin polarization in the
other electronic moments surrounding them. Following the work of others \cite
{Bobroff1}, we use decaying staggered spin susceptibility which we take to
be exponential, namely, 
\begin{equation}
\chi ^{^{\prime }}({\bf r})=(-1)^{n_{x}+n_{y}}\exp (-r/\xi )
\end{equation}
where ${\bf r}=n_{x}a\widehat{x}+n_{y}a\widehat{y}$ represents the position
of the neighbor Cu sites, ${\bf a}$ is the lattice vector, and $\xi $ is the
characteristic length scale. Because of this decay, at low frozen spin
concentration, large parts of the lattice are practically field free (expect
for nuclear moments). However, the important point is that no clear
distinction between magnetic and field free (superconducting) regions
exists. This situation is demonstrated in Fig.~\ref{Simulation}.

The muon polarization time evolution in this kind of field distribution is
numerically simulated. The interaction between the muon and all the other
moments is taken to be dipolar, and $\xi $ is taken to be 3 lattice
constants \cite{Kapitulnik1,NachumiPRL96}. The dashed line in Fig.~\ref{pol}
is a fit to the $T=350$ mK data, which yield $p=15$\% and maximum moment
size $\simeq 0.06\mu _{B}$ . As can be seen, the line fits the data very
well. However, as expected, the fit is sensitive to $p\xi ^{2}$ only, namely
the effective area of the magnetic islands, so longer $\xi $ would have
given smaller $p$. The field distributions and the polarization curve shown
in Fig.~\ref{model} were actually generated using the simulation. In (a) the
spin density is 15\% while in (b) the density is 35\%.
In panel (c) of Fig.~\ref{pol} we show the spin polarization for different
hole concentration, varying from 0\% to 35\% with the same $\xi =3$. The
resemblance between the simulation results as a function of $p$ and the muon
polarization as a function of temperature in panel (a) leads us to our
fourth conclusion that the freezing process is mostly a growth in the total
area of the frozen AF islands.

\section{Discussion}
\label{Discussion}

We now discuss the phase diagram presented in Fig.~\ref{clblco}. This
diagram is consistent with recent theories \cite{Assa} of hole pair boson
motion in an antiferromagnetic background. Those theories conclude that 
\begin{equation}
T_{c}\propto Jn_{s}  \label{Tctons}
\end{equation}
where $n_{s}$ is the superconducting carrier density, and $J$ is the
antiferromagnetic coupling energy \cite{UemuraNature}. Let us define $\Delta
p_{hl}=p_{hl}-p_{0}$ where $p_{hl}$ is the number of mobile holes and $p_{0}$
is the number of mobile holes at optimum.

\begin{figure}[tbp]
       \centerline{\epsfxsize=8.0cm \epsfbox{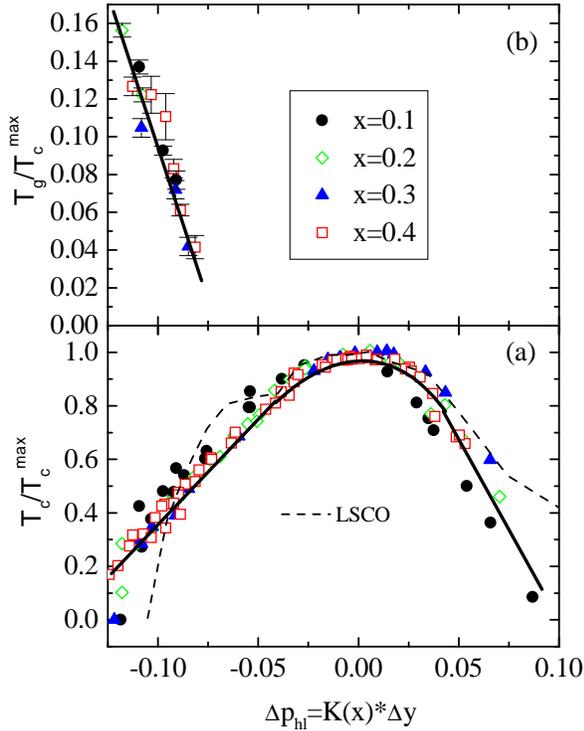}}              
\caption[]{$T_{c}/T_{c}^{max}$ and $T_{g}/T_{c}^{max}$ as a function of $K(x)\Delta y$
where $\Delta y=y-7.15$, and $K(x)$ is chosen so that all $T_{c}/T_{c}^{max}$
data sets collapse to a single curve, which resembles the LSCO curve.}
\label{scaling}
\end{figure}

 We assume that 
\begin{equation}
n_{s}(\Delta p_{hl})=\frac{1}{2}\left( p_{0}+\Delta p_{hl}\right)
\label{nstoDp}
\end{equation}
and write 
\begin{equation}
\Delta p_{hl}=K(x)\Delta y  \label{DptoDy}
\end{equation}
where $\Delta y=y-7.15$ is chemical doping measured from optimum, and $K$ is
a scaling parameter which relates $\Delta y$ to $\Delta p_{hl}$. Since there
is a linear dependence between $T_{g}$ and chemical doping (Fig.~\ref{Tg})
we predict that 
\begin{equation}
T_{g}\propto J(1-c_{g}n_{s}).  \label{Tgtons}
\end{equation}
>From Eq.~\ref{Tctons} $T_{c}^{\max }\propto Jn_{s}(0)$, therefore both $%
T_{c}/T_{c}^{\max }$ and $T_{g}/T_{c}^{\max }$ should be functions only of $%
\Delta p_{hl}$. This is demonstrated in Fig.~\ref{scaling}(a). We find $K(x)$
by making all $T_{c}/T_{c}^{\max }$ collapse onto one curve resembling the
curve of La$_{2-x}$Sr$_{x}$Cu$_{1}$O$_{4}$ \cite{PanCondMat02}, where the
exact doping is known. Using these values of $K(x)$ we also plot $%
T_{g}/T_{c}^{\max }$ as a function of $\Delta p_{hl}$ in Fig.~\ref{scaling}%
(b). Again all data sets collapse onto a single line described by $%
T_{g}/T_{c}^{\max }=-3.1(2)\Delta p_{hl}-0.21(2)$, or 
\begin{equation}
T_{g}=0.3T_{c}^{\max }(1-c_{g}n_{s})  \label{TgvsTcns}
\end{equation}
with $c_{g}=10.3$ when converting back to Eq.~\ref{Tgtons}. This indicates
that the same single energy scale $J$ controls both the superconducting and
magnetic transitions, and provides the exact $n_{s}$ and $T_{c}^{\max }$
dependence of $T_{g}$. 

\section{Conclusions}

We are now in a position to address the questions presented in the
introduction. The Uemura relations is adequately respected for our HTSC
``families''. We believe this is a result of the fact that there are no
structural changes between the different ``families''. The appearance of
spontaneous magnetic field in CLBLCO is a gradual process. As the
temperature is lowered microscopic regions of frozen moments appear in the
samples, and their area increases but the moments do not. In the ground
state the field profile is very different from that of a canonical spin
glass or any other standard magnet. It could only be generated by
microscopic intercalation of an abnormal number of zero field regions with
magnetic regions without a clear distinction between the two. Finally, and
most importantly, the phase diagram containing both $T_{c}$ and $T_{g}$
leads us to believe that these temperatures are determined by the same
energy scale.

\section{Acknowledgments}

We would like to thank the PSI and ISIS facilities for their kind
hospitality and continuing support of this project. We acknowledge very
helpful discussions with Assa Auerbach and Ehud Altman. This work was funded
by the Israeli Science Foundation and the EU-TMR program.


\begin{thebibliography}{99}
\bibitem{comment3}  The density of hole per plane is $p=-0.205+y/3$.
\bibitem{Comment1}  As long as $\gamma _{\mu }H<\nu $, where $\nu $ is the
fluctuation rate.
\bibitem{comment2}  A moment of $1\mu_{B}$ will induce a 1G field at a
distance of $\sim20 \AA$. This field is equivalent to the nuclear moments
background.
\end{thebibliography}
\end{document}